\documentclass[12pt]{article}
\usepackage{amsmath}

\usepackage{graphicx}
\usepackage{comment}
\renewcommand{\d}{{\rm d}}
\newcommand{\bx}{\mbox{\boldmath$x$}}
\newcommand{\bN}{\mbox{\boldmath$N$}}
\newcommand{\bxi}{\mbox{\boldmath$\xi$}}
\begin{document}

\title{On the Kerr metric in a synchronous reference frame
}

\author{V.M. Khatsymovsky \\
 {\em Budker Institute of Nuclear Physics} \\ {\em of Siberian Branch Russian Academy of Sciences} \\ {\em
 Novosibirsk,
 630090,
 Russia}
\\ {\em E-mail address: khatsym@gmail.com}}
\date{}
\maketitle
\begin{abstract}
The Kerr metric is considered in a synchronous frame of reference obtained by using proper time and initial conditions for particles that freely move along a certain set of trajectories as coordinates. Modifying these coordinates in a certain way (keeping their interpretation as initial values at large distances), we still have a synchronous frame and the direct analogue of the Lemaitre metric, the singularities of which are exhausted by the physical Kerr singularity (the singularity ring).
\end{abstract}


MSC classes: 83C15; 83C57

keywords: general relativity; Kerr spacetime; synchronous frame

\section{Introduction}

An exact solution of the vacuum Einstein equations describing gravitational field of a spinning mass was given by Kerr \cite{Kerr}. Depending on the problem under consideration, it is convenient to write the corresponding metric in one or another coordinate system; see, for example, \cite{Boyer} - \cite{Baines}. We issue from the Kerr metric in the Boyer-Lindquist coordinates \cite{Boyer},
\begin{eqnarray}\label{ds2-Boyer}                                           
\d s^2 & = & \left ( - 1 + \frac{r_g r}{ \rho^2 } \right ) \d t^2 + \frac{ \rho^2 }{ \triangle } \d r^2 + \rho^2 \d \theta^2 + \left ( r^2 + a^2 + a^2 \frac{r_g r}{ \rho^2 } \sin^2 \theta \right ) \sin^2 \theta \d \varphi^2 \nonumber \\ & & - 2 a \frac{r_g r}{ \rho^2 } \sin^2 \theta \d \varphi \d t , \\ & & \mbox{ where } \rho^2 = r^2 + a^2 \cos^2 \theta , ~ \triangle = r^2 - r_g r + a^2 . \nonumber
\end{eqnarray}

\noindent The contravariant metric tensor is
\begin{eqnarray}                                                            
\left \| g^{\lambda \mu } \right \| & = & \left [ \begin{array}{cccc} - \Sigma^2 / \rho^2 \triangle & 0 & 0 & - a r_g r / \rho^2 \triangle \\ 0 & \triangle / \rho^2 & 0 & 0 \\ 0 & 0 & 1 / \rho^2 & 0 \\ - a r_g r / \rho^2 \triangle & 0 & 0 & ( \triangle - a^2 \sin^2 \theta ) / \rho^2 \triangle \sin^2 \theta \end{array} \right ] , \\ & & \mbox{ where } \Sigma^2 = \left ( r^2 + a^2 \right )^2 - a^2 \triangle \sin^2 \theta . \nonumber
\end{eqnarray}

One of the important coordinate systems is the synchronous reference system. By fixing four components of the metric tensor, $g_{0 \lambda} = (-1, 0, 0, 0)$, the synchronous frame explicitly leaves us with six physically significant metric functions, the spatial metric. It is also the simplest case of the lapse-shift functions $(N, \bN) = (1, {\bf 0})$ in the Arnowitt-Deser-Misner formalism \cite{ADM1}, a way to transfer the physics of the phenomenon to the true canonical coordinate (spatial metric), leaving $(N, \bN) = {\rm const}$, which may be interesting in quantum theory.

In what follows, we will transform the metric (\ref{ds2-Boyer}) into a synchronous frame of reference, in which the coordinates are the proper time and initial conditions for a certain set of trajectories of freely moving particles, using a technique based on the Hamilton-Jacobi equation for a particle. The resulting metric has singularities in addition to the true ring Kerr singularity. Then we modify the definition of the new coordinates to some "asymptotic" form, so that their interpretation as initial values for the original coordinates (Boyer-Lindquist) will not necessarily be correct (it is correct at large distances), but the metric is simplified and only has the true ring Kerr singularity.

\section{Transformation to a synchronous frame}

A way of passing to a synchronous frame is to use the Hamilton-Jacobi equation for a particle with action $\tau$ (as mentioned, for example, in the textbook \cite{Landau}),
\begin{equation}                                                            
g^{\lambda \mu} \frac{\partial \tau}{\partial x^\lambda} \frac{\partial \tau}{\partial x^\mu} + 1 = 0 .
\end{equation}

\noindent For that, its solution is required,
\begin{equation}                                                            
\tau = f ( \bxi , \bx , t ) + A ( \bxi ) ,
\end{equation}

\noindent which depends on four constants $\bxi , A ( \bxi )$ as parameters, of which $A ( \bxi )$ is considered as an arbitrary function of $\bxi$. The equations of motion are
\begin{equation}                                                            
f_{, \xi_j} ( \bxi , \bx , t ) + A_{, \xi_j} ( \bxi ) = 0 .
\end{equation}

\noindent We consider the set of trajectories corresponding to a given fixed $\bxi$. We take $\tau$ as the new time coordinate and set $A ( \bxi ) = 0$ for the given $\bxi$ (this is tantamount to redefining $\tau$ by shifting). If at $\tau = 0$ the trajectory passing through $\bx , t$ has coordinates $\bx_1 , t_0 ( \bx_1 )$, then $\bx_1 , \tau$ are the new coordinates of the point $\bx , t$. We have
\begin{eqnarray}\label{x1,tau=x,t-general}                                  
\tau & = & f ( \bxi , \bx , t ) , \nonumber \\ f ( \bxi , \bx , t ) & = & 0 \Rightarrow t = t_0 ( \bx_1 ) , \nonumber \\ f_{, \xi_j} ( \bxi , \bx_1 , t_0 ( \bx_1 ) ) & = & f_{, \xi_j} ( \bxi , \bx , t ) .
\end{eqnarray}

The contravariant metric tensor in the new coordinates $\bx_1 , \tau$ has the components
\begin{eqnarray}                                                            
g^{\tau \tau} & = & g^{\lambda \mu} \frac{\partial f ( \bxi , \bx , t )}{\partial x^\lambda} \frac{\partial f ( \bxi , \bx , t )}{\partial x^\mu} = -1 , \nonumber \\ g^{x_1^j \tau} & = & \frac{\partial x_1^j}{ \partial x^\lambda} \frac{\partial \tau}{\partial x^\mu} g^{\lambda \mu} = \frac{\partial x_1^j}{ \partial f_{, \xi_k} ( \bxi , \bx_1 , t_0 ( \bx_1 ) )} \frac{\partial f_{, \xi_k} ( \bxi , \bx_1 , t_0 ( \bx_1 ) )}{ \partial x^\lambda} \frac{\partial \tau}{\partial x^\mu} g^{\lambda \mu} \nonumber \\ & = & \frac{\partial x_1^j}{ \partial f_{, \xi_k} ( \bxi , \bx_1 , t_0 ( \bx_1 ) )} \frac{\partial f_{, \xi_k} ( \bxi , \bx , t )}{ \partial x^\lambda} \frac{\partial f ( \bxi , \bx , t )}{\partial x^\mu} g^{\lambda \mu} \nonumber \\ & = & \frac{1}{2}  \frac{\partial x_1^j}{ \partial f_{, \xi_k} ( \bxi , \bx_1 , t_0 ( \bx_1 ) )} \frac{\partial }{\partial \xi_k } \left [ \frac{\partial f ( \bxi , \bx , t )}{ \partial x^\lambda} \frac{\partial f ( \bxi , \bx , t )}{\partial x^\mu} g^{\lambda \mu} ( \bx , t ) \right ] = 0 , \nonumber \\ g^{x_1^j x_1^k} & = & \frac{ \partial x_1^j }{ \partial x^\lambda } \frac{ \partial x_1^k }{ \partial x^\mu } g^{\lambda \mu} .
\end{eqnarray}

The Hamilton–Jacobi equation is completely separable in the Kerr geometry \cite{Carter} (see also the review \cite{Chandra}), and the completely separated solution is
\begin{eqnarray}                                                            
f ( \bxi , \bx , t ) = - E t + L \varphi + \int^r \frac{ \sqrt{ R} }{ \triangle } \d r + \int^\theta \sqrt{ \Theta } \d \theta , \nonumber \\ R = \left [ \left ( r^2 + a^2 \right ) E - a L \right ]^2 - \triangle \left [ Q + \left ( L - a E \right )^2 + r^2 \right ] , \nonumber \\ \Theta = Q + \left [ \left ( E^2 - 1 \right ) a^2 - \frac{ L^2 }{ \sin^2 \theta } \right ] \cos^2 \theta , \\ \mbox{ where } \triangle = r^2 - r_g r + a^2 . \nonumber
\end{eqnarray}

\noindent The constants of motion are energy $E$, angular momentum $L$ and a new constant $Q$. We choose $E = 1$ and $L = 0$, as for freely falling particles that start with zero velocity at infinity, with which the Lemaitre frame \cite{Lemaitre} can be associated in the Schwarzschild case. Then the values $Q < 0$ are not in the domain of definition. We set $Q = q^2$. Equations (\ref{x1,tau=x,t-general}) relating $\bx_1$, $\tau$ and $\bx$, $t$ at $\bxi = (E, L, q) = (1, 0, q)$ have the form
\begin{eqnarray}\label{x1,tau=x,t}                                          
\tau & = & - t + \int^r \frac{ \sqrt{ R} }{ \triangle } \d r + q \theta , \nonumber \\ f_{, E} : & & \int^{r_1} \frac{ r^2 + q^2}{\sqrt{ R }} \d r - q \theta_1 + \int^{\theta_1} \frac{ a^2 \cos^2 \theta}{ q } \d \theta \nonumber \\ & = & - t + \int^r \frac{ \left ( r^2 + a^2 \right )^2 - a^2 \triangle }{ \triangle \sqrt{ R }} \d r + \int^{\theta} \frac{ a^2 \cos^2 \theta}{ q } \d \theta , \nonumber \\ f_{, L} : & & - \int^{r_1} \frac{a r_g r}{ \triangle \sqrt{ R }} \d r + \varphi_1 = - \int^r \frac{a r_g r}{ \triangle \sqrt{ R }} \d r + \varphi , \nonumber \\ f_{, q} : & & - \int^{r_1} \frac{ q }{ \sqrt{ R }} \d r + \theta_1 = - \int^{r} \frac{ q }{ \sqrt{ R }} \d r + \theta .
\end{eqnarray}

\noindent At $q \neq 0$, the set of trajectories with $(E, L, q) = (1, 0, q)$ does not reach sufficiently small $r$ ($R < 0$ at these $r$). Therefore, we pass to the limit $q \to 0$. Equation (\ref{x1,tau=x,t}) obtained from $f_{, q}$ gives $\theta_1 - \theta = O ( q ) \to 0$, which allows finding $( \theta_1 - \theta ) / q$ in the equation from $f_{, E}$. The system (\ref{x1,tau=x,t}) gives a relation between the differentials of the coordinates, and we find the nontrivial $3 \times 3$ block of the contravariant components of the new metric, $g^{x_1^j x_1^k}$,
\begin{eqnarray}                                                           
& & \hspace{-3mm} \left \| g^{x_1^j x_1^k} \right \| = \\ & & \hspace{-3mm} \left [ \begin{array}{c|c|c} e_1^2 \left ( 1 + \frac{\rho^2 \triangle}{R} + \frac{h^2}{ \rho^2 } \right ) & e_1 \frac{h}{ \rho^2 } & e_1 \eta_1 \left [ 1 + \frac{\rho^2 \triangle}{R} \left ( 1 - \frac{ \eta }{ \eta_1 } \right ) + \frac{h^2}{ \rho^2 } \right ] \\ \hline e_1 \frac{h}{ \rho^2 } & \frac{ 1 }{ \rho^2 } & \eta_1 \frac{h}{ \rho^2 } \\ \hline e_1 \eta_1 \left [ 1 + \frac{\rho^2 \triangle}{R} \left ( 1 - \frac{ \eta }{ \eta_1 } \right ) + \frac{h^2}{ \rho^2 } \right ] & \eta_1 \frac{h}{ \rho^2 } & \eta_1^2 \left [ 1 + \frac{\rho^2 \triangle}{R} \left ( 1 - \frac{ \eta }{ \eta_1 } \right )^2 + \frac{h^2}{ \rho^2 } \right ] + \frac{ \rho^2 - r_g r}{ \rho^2 \triangle \sin^2 \theta } \end{array} \right ], \nonumber \\ & & \hspace{-3mm} \mbox{ where } e_1 = \frac{ \sqrt{ R_1 } }{ \rho_1^2 } , ~ \eta_1 = \frac{ a r_g r_1 }{ \rho_1^2 \triangle_1 } , ~ R = r_g r ( r^2 + a^2 ) , ~ h = a^2 \sin ( 2 \theta ) \int_r^{ r_1 } \frac{ \d r }{ \sqrt{ R } } , \nonumber
\end{eqnarray}

\noindent and the subscript 1 at functions means the replacement $r \to r_1$. The covariant nontrivial (spatial-spatial) components are as follows,
\begin{eqnarray}\label{ds2-synchronous}                                    
g_{r_1 r_1} & = & \frac{ \rho_1^4 }{ r_1 \left ( r_1^2 + a^2 \right )} \frac{1}{ \rho^2 } \left ( r - 2 \frac{ a^2 r_g r_1 }{ \rho_1^2 \triangle_1 } r \sin^2 \theta + \frac{ a^2 r_g r_1^2 }{ \rho_1^4 \triangle_1^2 } \Sigma^2 \sin^2 \theta \right ) , \nonumber \\ g_{r_1 \theta_1} & = & - a^2 \frac{ \rho_1^2 }{ \sqrt{ r_1 \left ( r_1^2 + a^2 \right ) } } \frac{ r \sin ( 2 \theta )}{ \rho^2 } \tilde{ I } \left ( 1 - \frac{ a^2 r_g r_1 }{ \rho_1^2 \triangle_1 } \sin^2 \theta \right ) , \nonumber \\ g_{r_1 \varphi_1} & = & a \sqrt{r_g} \frac{ \rho_1^2 }{ \sqrt{ r_1 \left ( r_1^2 + a^2 \right ) } } \frac{ \sin^2 \theta}{ \rho^2 } \left ( r - \frac{ r_1 }{ \rho_1^2 \triangle_1 } \Sigma^2 \right ) , \nonumber \\ g_{\theta_1 \theta_1} & = & \rho^2 + a^4 \frac{r \sin^2 ( 2 \theta )}{ \rho^2 } \tilde{ I }^2 , \nonumber \\ g_{\theta_1 \varphi_1} & = & - a^3 \sqrt{r_g } \frac{r \sin ( 2 \theta ) \sin^2 \theta }{ \rho^2 } \tilde{ I } , \nonumber \\ g_{\varphi_1 \varphi_1} & = & \frac{ \Sigma^2 }{ \rho^2 } \sin^2 \theta , \\ \mbox{ where } \tilde{ I } & = & \int_r^{r_1} \frac{ \d r }{ \sqrt{ r \left ( r^2 + a^2 \right )} } . \nonumber
\end{eqnarray}

\noindent The determinant is
\begin{equation}                                                           
\det \| g_{x_1^j x_1^k} \| = \frac{\rho_1^4 r \left ( r^2 + a^2 \right ) }{ r_1 \left ( r_1^2 + a^2 \right ) } \sin^2 \theta .
\end{equation}

\noindent The dependence of $r$ on $r_1$ and $\tau$ for a given $\theta$ ($= \theta_1$) is determined from the relation
\begin{equation}\label{tau=f(r1,r,theta)}                                  
\tau \sqrt{ r_g } = \int_r^{ r_1 } \frac{ r^2 + a^2 \cos^2 \theta }{ \sqrt{ r \left ( r^2 + a^2 \right ) } } \d r .
\end{equation}

\noindent Note that such a geodesic was obtained in \cite{Baines} in the Doran coordinates \cite{Doran} $\tau$, $r$, $\theta$.

\section{Asymptotic form of the transformation}

\noindent The metric obtained is singular if $r_1$, $\rho_1$ or $\triangle_1$ are equal to zero. However, these singularities are transient, since they are excluded if the moment in time is somewhat greater than $\tau = 0$. Namely, if
\begin{equation}                                                           
\tau > \tau_0, ~~~ \tau_0 \sqrt{ r_g } = \int_0^{ r_+ } \frac{ r^2 + a^2 }{ \sqrt{ r \left ( r^2 + a^2 \right ) } } \d r = \int_0^{ r_+ } \sqrt{ r + \frac{ a^2 }{ r }} \d r ,
\end{equation}

\noindent where $r_+$ is the larger of the roots of $\triangle = 0$ (horizon radius), then equation (\ref{tau=f(r1,r,theta)}) for any $\theta$ has a solution $r \geq 0$ only for $r_1 > r_+$, and we are left only with the physical ring singularity at $\rho^2 = 0$.

Moreover, we note that we can take both $r_1$ and $\tau$ arbitrarily large, while keeping $2 r_1^{3 / 2} / ( 3 \sqrt{ r_g } ) - \tau$ finite and hence other coordinates in the physical region of interest. Then the (covariant) metric tensor is simplified,
\begin{eqnarray}\label{ds2-Lemaitre-like}                                  
& & \hspace{-3mm} \left \| g_{x_1^j x_1^k} \right \| = \nonumber \\ & & \hspace{-3mm} \left [ \begin{array}{c|c|c} r_1 \frac{r}{\rho^2} & - \sqrt{r_1} a^2 \frac{r \sin ( 2 \theta )}{ \rho^2 } I & \sqrt{ r_1 } a \sqrt{ r_g } \frac{r \sin^2 \theta}{ \rho^2 } \\ \hline - \sqrt{r_1} a^2 \frac{r \sin ( 2 \theta )}{ \rho^2 } I & \rho^2 + a^4 \frac{ r \sin^2 ( 2 \theta ) }{ \rho^2 } I^2 & - a^3 \sqrt{ r_g } \frac{ r \sin ( 2 \theta ) \sin^2 \theta }{ \rho^2 } I \\ \hline \sqrt{ r_1 } a \sqrt{ r_g } \frac{r \sin^2 \theta}{ \rho^2 } & - a^3 \sqrt{ r_g } \frac{ r \sin ( 2 \theta ) \sin^2 \theta }{ \rho^2 } I & \left ( r^2 + a^2 + a^2 \frac{r_g r}{ \rho^2 } \sin^2 \theta \right ) \sin^2 \theta \end{array} \right ], \nonumber \\ & & \hspace{-3mm} \mbox{ where } I = \int_r^\infty \frac{ \d r }{ \sqrt{ r \left ( r^2 + a^2 \right )} } ,
\end{eqnarray}

\noindent and $r$ is regarded as a function of $\tau$, $r_1$, $\theta$ via
\begin{equation}\label{tau=f0(r1,r,theta)}                                 
\tau = \frac{ 2 }{ 3 } \frac{ r_1^{ 3 / 2 } - r^{ 3 / 2 } }{ \sqrt{ r_g } } + \int_r^\infty \left [ \frac{ r^2 + a^2 \cos^2 \theta }{ \sqrt{ r \left ( r^2 + a^2 \right )} } - \sqrt{ r } \right ] \frac{ \d r }{ \sqrt{ r_g } } .
\end{equation}

\noindent The contravariant metric tensor becomes even more simplified,
\begin{eqnarray}                                                           
& & \left \| g^{x_1^j x_1^k} \right \| = \left [ \begin{array}{c|c|c} \frac{ 1 }{ r_1 } \left [ r_g + \frac{\rho^2 \triangle}{ r ( r^2 + a^2 ) } + a^4 \frac{ \sin^2 ( 2 \theta )}{ \rho^2 } I^2 \right ] & \frac{ a^2 }{ \sqrt{ r_1 } } \frac{ \sin ( 2 \theta ) }{ \rho^2 } I & - \frac{ 1 }{ \sqrt{ r_1 } } \frac{ a \sqrt{ r_g }}{ r^2 + a^2 } \\ \hline \frac{ a^2 }{ \sqrt{ r_1 } } \frac{ \sin ( 2 \theta ) }{ \rho^2 } I & \frac{ 1 }{ \rho^2 } & 0 \\ \hline  - \frac{ 1 }{ \sqrt{ r_1 } } \frac{ a \sqrt{ r_g }}{ r^2 + a^2 } & 0 & \frac{ 1 }{ ( r^2 + a^2 ) \sin^2 \theta } \end{array} \right ].
\end{eqnarray}

Note that now we can consider $r_1$ and $\tau$ not necessarily arbitrarily large, and expression (\ref{ds2-Lemaitre-like}) for $ g_{x_1^j x_1^k}$ will still be accurate, only the interpretation of $r_1$ as the initial value for $r$ will not necessarily take place.

$2 r_1^{3 / 2} / ( 3 \sqrt{ r_g } )$ is an analogue of the Lemaitre \cite{Lemaitre} radial coordinate in the Schwarz\-schild case (expressing the Lemaitre metric in terms of $r_1$ instead of that coordinate was proposed in \cite{Stan}). At $a = 0$, we have the Lemaitre metric,
\begin{eqnarray}\label{Lem}                                                
& & \d s^2 = - \d \tau^2 + \frac{r_1}{r(r_1, \tau )} \d r^2_1 + r^2 (r_1 , \tau ) \d \Omega^2 , ~~~ r^{3/2} = r^{3/2}_1 - \frac{2}{3} \sqrt{r_g} \tau .
\end{eqnarray}

\begin{figure}[b]
\centerline{\includegraphics[width=10.0cm]{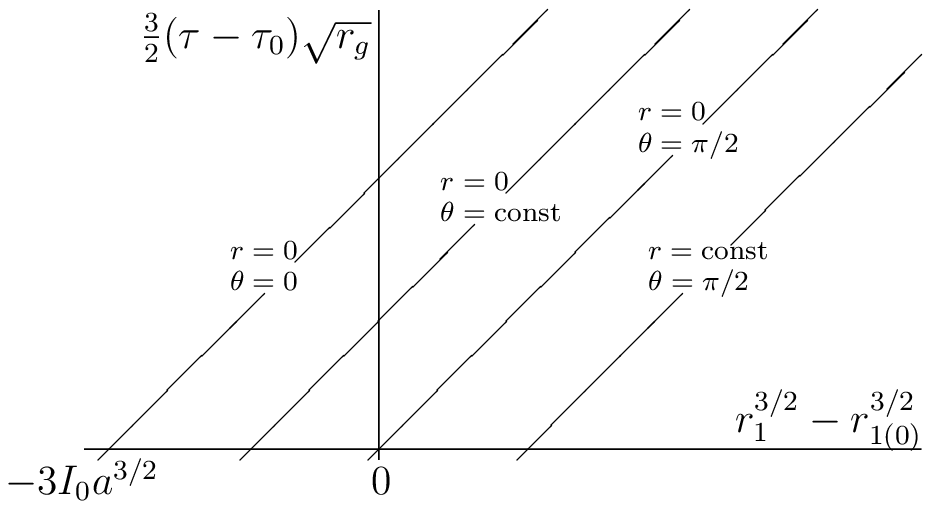}}
\caption{A section along $r_1$, $\tau$ passing through the singularity $r = 0$, $\theta = \pi / 2$. $I_0 = \int_0^\infty ( 1 + y^4 )^{ - 1 / 2 } \d y = [ \Gamma ( 1 / 4 ) ]^2 / ( 4 \sqrt{ \pi } )$. \label{f4}}
\end{figure}

In Fig.~\ref{f4}, a section along $r_1$, $\tau$ passing through the singularity $r = 0$, $\theta = \pi / 2$ is shown.
There are two regions, $r = 0$, $0 \leq \theta < \pi / 2$ and $r > 0$, $\theta = \pi / 2$ (the inside and outside of the singularity ring in the equatorial plane), on either side of the singularity line $r = 0$, $\theta = \pi / 2$, compared to the Lemaitre metric, for which there is only $r > 0$ on one side.

From (\ref{tau=f0(r1,r,theta)}), we can find for the differentials
\begin{equation}                                                           
\sqrt{ r_1 } \d r_1 = \sqrt{ r_g } \d \tau + \frac{ \rho^2 }{ \sqrt{ r ( r^2 + a^2 ) } } \d r + a^2 I \sin ( 2 \theta ) \d \theta .
\end{equation}

\noindent If we exclude $r_1$ in favor of $r$, we should obtain an analogue of the Painlev\'{e}-Gullstrand metric \cite{Painleve,Gullstrand} for the Schwarzschild geometry, at least in the way of obtaining. This turns out to be just the Doran metric \cite{Doran},
\begin{eqnarray}\label{ds2-Painleve-Gullstrand-like}                       
& & \d s^2 = \left ( - 1 + \frac{r_g r}{\rho^2} \right ) \d \tau^2 + \frac{\rho^2}{r^2 + a^2} \d r^2 + \rho^2 \d \theta^2 + 2 \sqrt{ \frac{ r_g r }{ r^2 + a^2 } } \d \tau \d r \nonumber \\ & & + 2 a \sqrt{ \frac{ r_g r }{ r^2 + a^2 } } \sin^2 \theta \d r \d \varphi_1 + 2 a \frac{ r_g r }{ \rho^2 } \sin^2 \theta \d \tau \d \varphi_1 \nonumber \\ & & + \left ( r^2 + a^2 + a^2 \frac{r_g r}{ \rho^2 } \sin^2 \theta \right ) \sin^2 \theta \d \varphi_1^2 .
\end{eqnarray}

\noindent Vice versa, substituting $r = r (\tau, r_1, \theta )$, we can obtain a synchronous metric from the Doran one. In the present consideration, an asymptotic (at $\tau$, $r_1$ large) connection between these new coordinates and the coordinates bound to the set of freely moving particles is shown.

\section{Conclusion}

Thus, we have made a binding of coordinates to the set of timelike geodesics which represent the motion of freely falling particles with $(E, L, Q) = (1, 0, 0)$. This gives a synchronous frame. The following two points seem to be interesting. First, we cannot set the Carter’s constant $Q$ to be zero from the beginning, but we should carefully tend to zero, starting with small positive values. Second, the metric components (\ref{ds2-synchronous}) in the obtained frame of reference has singularities additional to the true Kerr singularity. However, these singularities are absent if the coordinates of the proper time $\tau$ and the initial radial coordinate $r_1$ are chosen large and at the same time corresponding to the points of interest in the original coordinates (Boyer-Lindquist). Moreover, we can take the asymptotic (at large $\tau$, $r_1$) form (\ref{tau=f0(r1,r,theta)}) of the transformation (\ref{tau=f(r1,r,theta)}) from $r$ to $r_1$ and get the metric in a slightly modified synchronous frame of reference (\ref{ds2-Lemaitre-like}). This metric is singular only in the true Kerr singularity (the singularity ring) and is the direct analogue of the Lemaitre metric in the Schwarzschild case, only the interpretation of $r_1$ as the initial value for $r$ will not necessarily take place.

\section*{Acknowledgments}

The present work was supported by the Ministry of Education and Science of the Russian Federation.

\end{document}